\documentclass[twocolumn,aps,prl,floatfix,showpacs]{revtex4}

\usepackage{graphicx}
\usepackage{dcolumn}
\usepackage{bm}

\begin{document}

\title{Single-Dirac-Cone topological surface states in TlBiSe$_2$ class of Topological Insulators}
\author{Hsin Lin$^1$, R.S. Markiewicz$^1$, L.A. Wray$^{2,3}$, L. Fu$^4$,
M.Z. Hasan$^{2,3}$, A. Bansil$^1$}

\affiliation{$^1$Department of Physics, Northeastern University, Boston,
Massachusetts 02115, USA}

\affiliation{$^2$Joseph Henry Laboratories of Physics, Princeton
University, Princeton, New Jersey 08544, USA}

\affiliation{$^3$Princeton Center for Complex Materials, Princeton
University, Princeton, New Jersey 08544, USA}

\affiliation{$^4$Department of Physics, Harvard University, Cambridge,
Massachusetts 02138, USA}

\begin{abstract}

We have investigated several strong spin-orbit coupling ternary
chalcogenides related to the (Pb,Sn)Te series of compounds. Our
first-principles calculations predict the low temperature rhombohedral
ordered phase in TlBiTe$_2$, TlBiSe$_2$, and TlSbX$_2$ (X=Te, Se, S) to be topologically Z$_2$ = -1 nontrivial. We identify the specific surface termination that realizes the single Dirac cone through first-principles surface state computations. This termination minimizes effects of dangling bonds making it favorable for photoemission (ARPES) experiments. Our analysis predicts that thin films of these materials would harbor novel 2D quantum spin Hall states, and support odd-parity topological superconductivity. For a related work also see arXiv:1003.2615v1. Experimental ARPES results will be published elsewhere.

\end{abstract}

\maketitle

Topological insulators are a recently discovered new phase of quantum
matter \cite{moore, hasankane, zhang}.  The search for topological
insulators in real materials has benefitted from the fruitful interplay
between topological band theory of Kane and Mele (2005) and realistic band structure calculations
\cite{HgTe, fukane, bisezhang, bisb, bisb2, bise, bite, bitedavid,
davidprl}.  As a result, Bi$_x$Sb$_{1-x}$, Bi$_2$Se$_3$ and Bi$_2$Te$_3$
have been experimentally realized as three-dimensional topological
insulators\cite{bisb, bise, bisb2, bite, bitedavid, davidprl}. Recently,
this search has been extended to ternary compounds \cite{heuslerhasan,
heuslerzhang}. Here, we report first-principles band calculations of
Tl-based III-V-VI$_2$ ternary chalcogenide series, and compare the results
to those of the related (Pb,Sn)Te series studied previously in connection
with Dirac fermion physics in the 1980s \cite{fradkin}. The low
temperature rhombohedral ordered phase in TlBiTe$_2$, TlBiSe$_2$, and
TlSbX$_2$ (X=Te, Se, S) is predicted to be topologically nontrivial.
Moreover, we have carried out first-principles slab computations in order
to identify the specific surface termination which gives rise to the
simple Dirac-cone surface band for ARPES measurements. An analysis of the symmetry of states
indicates that thin films of the present materials would support
2D-quantum spin Hall states.

\begin{figure}
\includegraphics[width=0.5\textwidth]{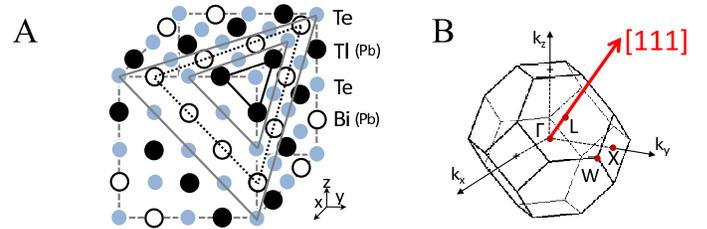}
\caption{\label{fig:sketch} (Color Online)
(A) Crystal structure of
idealized FCC TlBiTe$_2$. Tl surface
is emphasized by black lines, Bi surface by dotted lines, and the Te
surface by gray lines. (B) FCC Brillouin zone. Thick red arrow marks the
111 rhombohedral distortion direction.
}
\end{figure}

Designing new topological insulators involves modifying atomic structure
or doping to shift band orders out of the natural sequence. Consider, for
example, the well-known case of (Pb,Sn)Te with rocksalt structure.  The
end phase PbTe with face-centered cubic (FCC) lattice is topologically
trivial. In contrast, SnTe has band inversions at four equivalent
$L$-points where parities of conduction and valence bands are switched
(Fig.~2E). Since this inversion occurs at an even number of points in the
Brillouin zone, SnTe is also a topologically trivial band insulator. Fu
and Kane\cite{fukane} proposed that a rhombohedral distortion along a
particular 111 direction can induce (Pb,Sn)Te into a strong topological
phase because then the band inversion occurs only at the $L$-point along
the 111 direction which is distinguished from the other three $L$-points
(Fig.~\ref{fig:sketch}B).

\begin{figure}
\includegraphics[width=0.5\textwidth]{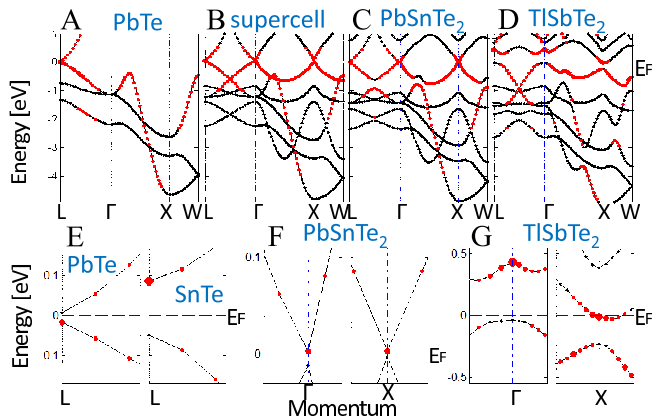}
\caption{ (Color Online)
(A)-(D) Band structures of
PbTe; PbTe in a supercell rhombohedral structure with 4 atoms/unit cell;
PbSnTe$_2$; and TlSbTe$_2$ with FCC atomic positions.
FCC symbols for points in the Brillouin
zone are used as in Fig.~1(B) for ease of comparison. Size of
red dots denotes the probability of s-orbital occupation on the $M$
atoms. (E) Band inversion at L for SnTe relative to PbTe.
(F) and (G) highlight band inversion at the $\Gamma$- and X-points for
PbSnTe$_2$ (C) and TlSbTe$_2$ (D), respectively.
}
\end{figure}

The Tl-based III-V-VI$_2$ ternary chalcogenides $MM'X_2$ are, or can
be, approximately viewed as a rhombohedral structure with space group
$R\bar{3}m$\cite{ThalliumPRB} with a center of inversion
symmetry. The unit cell contains four atoms: $M$ = Tl, $M'$ = Bi or
Sb, and $X$ = Te, Se, or S atoms, which occupy the Wyckoff 3a, 3b, and 6c
positions, respectively. If $M$ = $M'$ and the corresponding hexagonal
lattice constants a and c satisfy the relation c = 2$\sqrt{6}$a,
a rocksalt FCC structure is restored
with a$'$ = $\sqrt{2}$a with only two atoms/unit
cell. The rhombohedral lattice is embedded in a 2$\times$2 supercell of
the FCC
lattice as shown in Fig.~\ref{fig:sketch}A. This relation between
rhombohedral and FCC lattices is analogous to the type II antiferromagnet
NiO with $M$ = Ni spin up, $M'$ = Ni spin down, and $X$ = O. In this
sense,
Tl-based III-V-VI$_2$ ternary chalcogenides constitute
an offshoot of the IV-VI semiconductors and can be called
pseudo IV-VI semiconductors\cite{pseudoPbTe, tlbite2structure, Tlmodel,
TlBiTe2inversion,cooling,TlSbTe2struct}.  Taking the example of
TlBiTe$_2$, because
Tl and Bi precede and follow Pb in the periodic table, TlBiTe$_2$
resembles PbTe where $M$ = $M'$ = Pb. Therefore, by distorting the PbTe
crystal along a 111 direction and replacing Pb by Bi and Tl
alternating along that 111 direction, one obtains TlBiTe$_2$ with
rhombohedral structure.
In the ordered crystalline phase, the X atom moves away from the FCC position.

Since Tl-based III-V-VI$_2$ ternary chalcogenides provide a rhombohedral
version of (Pb,Sn)Te, one may expect these compounds to display
topologically non-trivial phases.  We have explored this possibility by
performing
first-principles
calculations \cite{wien2k} within
the framework of the density functional theory (DFT) using the generalized
gradient approximation (GGA)\cite{PBE96}.
Spin orbital coupling (SOC) was included
as a second variational step.
The lattice
constants were taken from the optimized values given in table I of
Ref.~\onlinecite{ThalliumPRB}.

To better explain the relationship between (Pb,Sn)Te and the Tl-based
III-V-VI$_2$ ternary chalcogenides, Fig.~2 shows the evolution of
bands for a sequence of structures going from FCC PbTe
(Fig.~2A) to TlSbTe$_2$ (Fig.~2D). For
FCC PbTe, the top of the valence bands is located at the
L-points with a small gap. When the same calculation is repeated by
assuming a rhombohedral unit cell with two chemical formula
units, doubled along a 111 axis, we obtain the folded bands of
Fig.~2B. In
particular, the top of the valence band at one L-point is folded to the
$\Gamma$-point, and the other three $L$-points are folded to
three equivalent $X$ points.
Since no distortions are introduced yet, the bands
do not interact with the folded bands.  In the next
step (Fig.~2C), we replace one of the Pb atoms by an Sn
atom, resulting in alternating Pb and Sn planes along the selected 111
direction. The original and folded bands now interact, inducing new band
gaps and avoided crossings for the hypothetical ternary compound
PbSnTe$_2$. Otherwise, the band structure is superficially
similar to that in Fig.~2B. The
conduction and valence bands now become inverted at high symmetry points.
The band inversion effect can be
monitored through the probability of s-orbital occupation
(red dots)\cite{swave}
on the $M$ atom
(here $M$ = Pb) at the inversion center.
Comparing the band sequence with red dots for PbTe and PbSnTe$_2$, we see
that band inversion indeed occurs at both $\Gamma$ and
$X$-points (Fig.~2F).  Due to an even number of band inversions,
PbSnTe$_2$ is also topologically trivial similar to SnTe. But, since
PbSnTe$_2$ only has rhombohedral instead of cubic symmetry, its band gaps
at the $\Gamma$ and $X$-points become unequal. The band gaps at $X$ points
are smaller, which indicates that band inversion from PbTe to
PbSnTe$_2$ occurs first at $\Gamma$ and then at $X$.  This suggests that
by introducing a small rhombohedral lattice distortion or by replacing Pb
and Sn with other atoms, the second band inversion associated
with the three $X$-points might be removed, leaving only a single band
inversion relative to PbTe (in supercell) at $\Gamma$, which
would lead to a topologically non-trivial band structure.
This leads us to replace Pb (Sn) by Tl (Sb) and introduce rhombohedral
lattice
distortion to obtain the bands in Fig.~2D, which resemble those of
(Pb,Sn)Te.  The rhombohedral lattice
distortion enhances all band gaps, especially at $\Gamma$, and shifts the
band edges at the $X$-points.
The bottom of the
conduction band at the $X$-point overlaps the top of the valence band at a
low symmetry point (not shown), leading to a semi-metal ground state.
Nevertheless, since a direct
band gap between the conduction and valence bands exists throughout
the Brillouin zone, the $Z_2$ topology of the valence bands is well
defined.  To determine the band topology, one should focus on the band
inversion relative to supercell PbTe at $\Gamma$ and three
$X$-points.  The s-orbital occupation in Fig.~2D shows that band
inversion occurs at all four points as in PbSnTe$_2$.
Hence, in FCC structure, TlSbTe$_2$ and indeed all six Tl-based
III-V-VI2 ternary
chalcogenides presented here, would be topologically trivial semi-metals.

\begin{figure}
\includegraphics[width=0.5\textwidth]{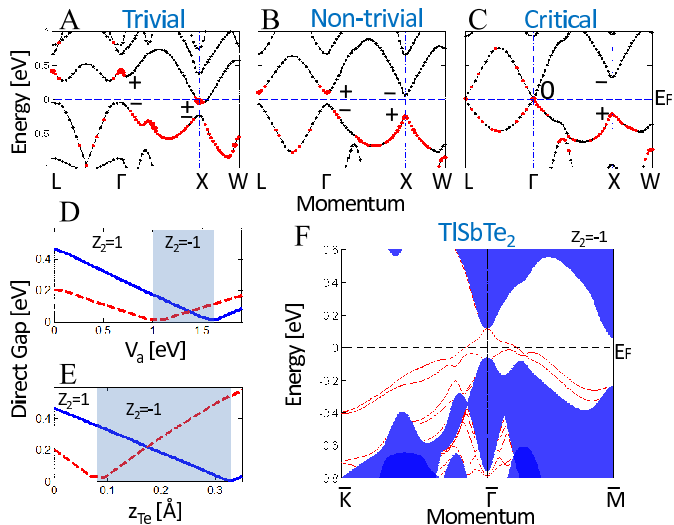}
\caption{\label{fig:bulkbands} (Color Online)
(A)-(C) show the bulk band structure of TlSbTe$_2$ for
$z_{Te}$ = 0, 0.21, and 0.33 $\AA$, respectively.
FCC notation for the high-symmetry points is used as in Fig.~1B.
Panels (D) and (E) show the direct gaps at $\Gamma$ (blue solid line)
and $X$ (red dashed line)
as a function of $V_a$ and $z_{Te}$, respectively.
The gray
area highlights features related to topological non-triviality. Panel (F)
shows the surface state dispersions at optimized value of $z_{Te}$ in red lines.
Projected bulk bands are
shown as blue areas.
}
\end{figure}

However, experiments \cite{tlbite2structure,cooling,TlSbTe2struct} find
that the X atom moves away from the FCC position
towards the $M'$ layer in the ordered crystalline phase.
We define $z_{X}$ to be the distance
away from the FCC position for the X atom.
Fig.~3 gives band structures of TlSbTe2 for the DFT-GGA
total-energy optimized $z_{Te}$= 0.21 $\AA$ value
as well as for two shifted values of $z_{Te}$ to illustrate
the sensitivity of band structure to the value of $z_{Te}$.
The optimized value is in good agreement with experiments\cite{TlSbTe2struct}.
For $z_{Te}$ = 0, the band
parities at $\Gamma$ and $M$ have the same ordering as that in SnTe
(Fig.~2E) or PbSnTe$_2$ (Fig.~2F), and the material is topologically
trivial. For $z_{Te}$ = 0.21 $\AA$, the band parity at $X$-point changes
and the $Z_2$ index picks up a factor of -1. In addition, the gap between
the conduction and valence bands becomes large enough to make the compound
an insulator. Thus, TlSbTe$_2$ is predicted to be a topological insulator
for the optimized Te position. For $z_{Te}$ = 0.33 $\AA$, the gap
at $\Gamma$ is close to zero.
As Te atoms move further away, the material once again becomes
topologically trivial
with $Z_2$ = 1.
The range of
$z_{Te}$ values which yield a non-trivial phase is shown by
the gray area in Fig.~3E. This range can be ascertained by monitoring the
gap sizes at $\Gamma$ and $X$. At the critical point where the band
inverts, the gap goes to zero. We have repeated the preceding procedure
for other Tl-compounds and find that all these compounds become
topologically nontrivial at the optimal X-atom
position except for TlBiS$_2$, which is predicted to be a topologically
trivial band insulator.\cite{footzhang,thalliumzhang}

In order to test the robustness of our results to effects of
band gap corrections or alloying, we have considered effects of
adding an orbital-dependent potential into the Hamiltonian.
For Tl-based III-V-VI$_2$ ternary chalcogenides, the orbital character of
the conduction (valence) band is p-type for $M'$ ($X$) atoms.
We specifically add orbital-dependent
potentials
$V_a$ to all three p-orbitals of $M'$,
which moves the conduction band upwards. As an
example,
we start with a topologically trivial phase in TlSbTe$_2$ with $z_{Te}$ =0.
With increasing $V_a$, as Fig.~3D shows, the gap at
$X$ vanishes at $V_a$=1 eV and band inversion occurs.
This transition occurs for the $\Gamma$ point at a
larger $V_a$ value of 1.6 eV. Between these two critical points, bands
are inverted only at the $\Gamma$-point relative to PbTe and the system
becomes topologically non-trivial (shaded area in Fig.~3D).
We have found a similar
behavior
in other compounds of this family. Despite uncertainties inherent in
first-principles
computations, our calculations establish band inversion at
$\Gamma$ and $X$-points (relative to PbTe) as the decisive factor of their
topological class.

A non-trivial band topology will generate metallic surface states which
are the hallmark of topological insulators. For the 111 surface in
Tl-based III-V-VI$_2$ ternary chalcogenides, the atoms are sequenced as
-$M$-$X$-$M'$-$X$- and there are four possible surface terminations (see
Fig.~1A).  Unlike Bi$_2$Se$_3$ and Bi$_2$Te$_3$,
the bonding between the layers in TlM'X$_2$ is not weak. As a result, the
topologically protected
surface states may coexist with non-topological ones arising from dangling
bonds, leading to a complicated surface spectrum. We have carried out
extensive slab calculations on all possible surface terminations for all
six compounds to search for a simple surface spectrum. We find that the
number of trivial dangling-bond surface states is minimized for a
termination with $X$ atoms exposed and $M'$ atoms beneath the surface,
since fewer bonds are broken in forming this surface, making this
the likely candiate for the naturally occuring surface. Fig.~3F shows
the surface band structure of TlSbTe$_2$ with optimized atomic positions
in a topologically non-trivial phase.
The calculation is based on a slab with 47 atomic layers.
The Dirac point is at the edge of the conduction band at
$\bar{\Gamma}$.
Between $\bar{\Gamma}$ and $\bar{M}$, there is only one surface band
(red lines)
connecting conduction and valence bands. This is unambiguous evidence for
a non-trivial topology.

\begin{figure}
\includegraphics[width=0.5\textwidth]{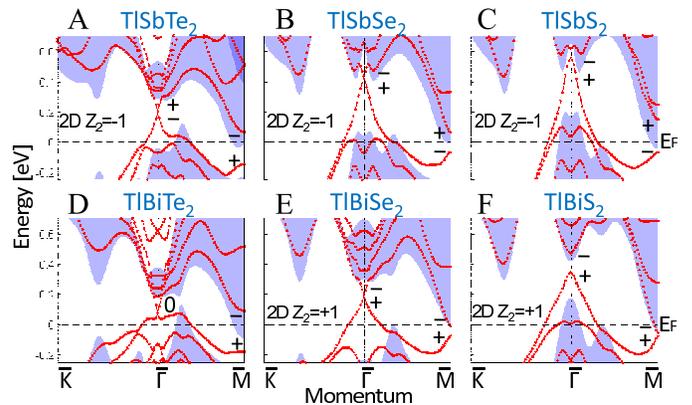}
\caption{\label{fig:HgTe} (Color Online)
Band structures based on
thin slab calculations for TlSbTe$_2$ (A), TlSbSe$_2$ (B), TlSbS$_2$ (C),
TlBiTe$_2$ (D), TlBiSe$_2$ (E), and TlBiS$_2$ (F) with $z_X$ = 0,
but without
bandgap corrections. Blue area marks projected bulk bands.
Red dots are 2D bands for the thin slab.
Parity analysis of states at $\bar{\Gamma}$
and $\bar{M}$ indicates that thin films in panels (A)-(C) would harbor
topologically nontrivial 2D quantum spin Hall states.}
\end{figure}

Even if the compounds are in a topologically trivial phase, their surface
states may have Dirac-cone dispersions due to band inversion.  In
Fig.~4, we show the calculated surface dispersions with $z_{Te}$ = 0
without bandgap corrections for a 23 layer slab where the $X$ atoms
are exposed at both sides of the slab.  The slab thickness is about
50{\AA}.  Dirac cones with small gaps are obtained at $\bar{\Gamma}$ and
$\bar{M}$ due to the projection of $\Gamma$ and $X$ points to the surface
Brillouin zone.  The Dirac cones have smaller gaps at $\bar{\Gamma}$ in
TlSbTe$_2$ (A), TlSbSe$_2$ (B), TlSbS$_2$ (C), TlBiTe$_2$ (D), and
TlBiSe$_2$ (E), and at $\bar{X}$ in TlBiS$_2$ (F).  Consider TlBiTe$_2$
as an example.  The gap at $\Gamma$ decreases rapidly as the
slab thickness increases, indicating that a gapless Dirac band exists at
$\Gamma$ on the surface in the infinite thickness limit.  Note however
that the existence of a surface Dirac band at $\Gamma$ by itself does not
reveal the band topology, and the surface spectrum near the
$\bar{M}$-points must also be considered. Hence a $k\cdot p$ model around
$\Gamma$ is not adequate for these compounds. In our slab calculations,
surface states near $\bar{M}$ have strong finite size effects arising from
interactions between the two end surfaces of the slab.
This sensitivity to finite size effects could be utilized to
design thin film samples which have quantum spin Hall states. In these
films, the Dirac points are all gapped, as in Fig.~4, and we can calculate
the 2D topological index $Z_2$ by analyzing only the points
$\bar{\Gamma}$ and $\bar{M}$.  From wavefunction parity analysis at these
points, illustrated in Fig.~4, we obtain $Z_2$ = -1 for TlSbTe$_2$,
TlSbSe$_2$, TlSbS$_2$. Hence these compounds are topologically non-trivial
quantum spin Hall systems. In the case of TlBiTe$_2$, the upper and lower
Dirac cones are almost degenerate at $\bar{\Gamma}$.

Interestingly, AgBiTe$_2$, AgBiSe$_2$, and AgInSe$_2$ all possess a
crystal structure similar to the present compounds, but with $M$ = Ag.
However, our first-principles bulk and surface state computations on
these Ag-compounds indicate lack of band inversions. In particular, the
surface and bulk bands display similar dispersions and there is no
evidence for the presence of a protected Dirac cone surface band or other
signatures of topologically interesting behavior.

Apart from the fact that Tl-compounds are usually semi-metallic or very
weakly semiconducting, they are topologically similar to the
Bi$_2$Se$_3$ series discovered previously. Recently, it was found
that upon Cu doping, Bi$_2$Se$_3$ becomes a superconductor with a T$_c$ =
3.8K and exhibits unconventional band topology \cite{hor, wray, toposc}.
The doped topological states in the Tl-based compounds thus are also a
likely
candidate for finding odd-parity topological superconductor similar to the
Cu$_x$Bi$_2$Se$_3$ series \cite{hor, wray, toposc}.

In conclusion, we have shown that the Tl-based III-V-VI$_2$ ternary
chalcogenides are highly favorable for supporting Z$_2$ =-1 topological
states.
Their topological nature is driven by the band inversions
at {\it both} $\Gamma$ and $X$-points.
We identify the optimal surface termination for
single Dirac cone surface states and predict that thin films could harbor
2D topologically non-trivial quantum spin Hall states. Our study indicates thus that the present Tl-based compounds constitute a fertile ground for observing and engineering topologically interesting phases of quantum matter. 

We acknowledge discussions with R.J. Cava and B. Barbiellini. The work at Northeastern and Princeton is supported by the Basic Energy Sciences, US Department of Energy (DE-FG02-07ER46352, DE-FG-02-05ER46200 and AC03-76SF00098), and benefited from the allocation of supercomputer time at NERSC and Northeastern University's Advanced Scientific Computation Center (ASCC). Support of A. P. Sloan Foundation (LAW and MZH at Princeton) and Harvard Society of Fellows (LF) is acknowledged.


\begin{thebibliography}{99}
\bibitem{swave}
Since the probability of s-orbital occupation (denoted by
the size of red
dots) must vanish for an odd-parity band at
high symmetry points, we can use this as a measure of band parity.
\end{thebibliography}
\end{document}